\documentclass[
showpacs,amsmath,amssymb]{revtex4}
\usepackage{amssymb}
\usepackage[dvips]{graphicx}
\begin{document}
\title{Theory of Extrinsic and Intrinsic Tunnelling in Cuprate Superconductors}
\author{J. Beanland and A. S. Alexandrov}
\affiliation{Department of Physics, Loughborough University, Loughborough LE11 3TU, United Kingdom\\}

\begin{abstract}
A theory capable of explaining intrinsic and extrinsic tunnelling conductance in underdoped cuprates has been devised that accounts for the existence of two energy scales, their temperature and doping dependencies. The asymmetry and inhomogeneity seen in extrinsic (normal metal - superconductor (NS)) tunnelling and the normal-state gapped intrinsic (SS) conductance is explained, as well as the superconducting gap and normal state pseudogap and the temperature dependence of the full gap.
\end{abstract}


\maketitle

\section{Introduction}
Since the discovery of high-$T_c$ superconductivity in 1986 by Bednorz and M\"uller \cite{muller}, there has been a huge theoretical effort to understand the mechanism behind it.
A lanthanum barium copper oxide was the first compound displaying this phenomenon \cite{muller}, now we know there are many compounds displaying high temperature superconductivity that involve copper and oxygen, these make up the cuprate family. Cuprates are distinguishable from conventional metallic superconductors by originating from the doping of the parent charge-transfer insulators. The superconducting parts are weakly coupled two dimensional doped layers held together by the parent lattice. Cuprates have unique properties, as well as their high-$T_c$ they also have two energy scales, or gaps: The BCS-like ``superconducting" gap (SG) present in cuprates and other related compounds develops below the superconducting critical temperature and can be seen by extrinsic and intrinsic tunnelling experiments as well as high-resolution angle-resolved photoemission (ARPES) experiments; there also exists another energy gap, the ``pseudogap" (PG) which is a large anomalous gap that exists well above $T_c$.
The PG phenomena was first observed in spin responses \cite{imai,imaietal} and using scanning tunnelling spectroscopy \cite{geerk} in underdoped YBa$_2$Cu$_3$O$_{7-\delta}$. Shortly after, the same gap was observed through infrared measurements \cite{schlesinger}, many experiments have since exhibited this PG.
The first explanation of the gap was offered in the form of real-space preformed hole pairs \cite{aleray} called small bipolarons which are bound together by a strong electron-phonon interaction (EPI). Since then many theoretical explanations have been proposed for the origin of the PG which can roughly be divided into two groups.
The first of these groups argues that the PG originates from some order, either static or fluctuating. The second understands the PG is the precursor of the SG, and reflects pair fluctuations above $T_c$.
Some of the theories from the first group see the superconducting state of cuprates as being the result of a doped Mott-insulator (for example \cite{fischer}, see \cite{leeetal} for a review). In his resonating valence bond (RVB) theory, Anderson focuses on the ground state and low lying excitations, the origin of the PG is seen as the spin gap associated with the breaking of RVB singlets \cite{anderson2004}. It has been suggested that adding impurities to (or doping) cuprates could weaken the order parameter (for example this order parameter could be antiferromagnetic spin fluctuations \cite{kampf}) and thus be the cause of the PG. Some believe the PG could be the result of SU2 rotations (for example \cite{wenlee}), in the underdoped region which connect fluctuations of staggered flux states and $d$-wave superconductivity. Chakravarty et al (2001) \cite{chak} proposed a static orbital current state called a $d$-density wave was the origin of the PG based on phenomenological grounds.
It has been argued that the PG is a consequence of a spin density wave (SDW) or charge density wave (CDW) state \cite{Varlamov}, or an interplay between the two \cite{pradhan}. It has also been suggested that the PG could be the result of inhomogeneous charge distributions containing hole-rich and hole-poor domains \cite{Melloetal, emery}, or the cause of the SG and PG could be the inter-band pairing of an itinerant band and defect states \cite{kristoffel}.

The second group bases the understanding of the PG and high-$T_c$ superconductivity on pairing interactions, preformed Cooper pairs have been suggested \cite{emkiv} where the pairing is not in real-space but instead in momentum-space. It has however been implied that the short coherence length of cuprate superconductors suggests they lie somewhere between the BCS limit of very large momentum-space pairs and the opposite case of small real-space pairs undergoing a Bose-Einstein condensation (BEC) \cite{chen}.
The BCS-BEC crossover has been studied in detail, for example in Ref.\cite{ranninger} a superfluid state is approached in a system of localised bosons (tightly bound electron pairs) in contact with a reservoir of itinerant fermions (electrons), it is assumed the spontaneous decay and recombination between the two species causes superconductivity and the PG is a consequence of this, opening up in the fermionic density of states (DOS).
Attractive Hubbard models have been considered as the origin of superconductivity and the PG. For example, studies of the normal-state of the two dimensional attractive Hubbard model have been carried out using  quantum Monte Carlo (QMC) calculations \cite{randeria}, also the excited and ground state properties of the two dimensional attractive Hubbard model have been studied using the conserving, self consistent T-matrix formalism in the intermediate coupling regime and at low electron concentration \cite{micnas}.
Other approaches emphasize the weak phase stiffness in underdoped cuprates which is a result of low superfluid density or superconducting carrier density and it leads to a suppression of $T_c$ by phase fluctuations, this means the underdoped cuprates are characterised by a relatively small phase stiffness and poor screening \cite{emkiv}.
A diagrammatic theory of the one-band Hubbard model was proposed, where the temperature of the onset of the PG is related to the scattering rate \cite{schmalian}. The effects of classical phase transitions on the quasiparticle spectra were contemplated in underdoped cuprates in the PG regime above $T_c$ by taking into account mean-field $d$-wave quasiparticles that are semiclassically coupled to supercurrents induced by fluctuating unbound vortex-antivortex pairs \cite{framil}. Another idea with incoherent $d$-wave quasiparticles suggests that when the phase-coherence length exceeds the Cooper pair size, a PG appears \cite{lammert}, the phase fluctuations of a $d_{x^2-y^2}$  pairing gap in a two dimensional BCS-like Hamiltonian approach is thought to be the origin of the PG \cite{eckl}. A phenomenological theory was produced that allowed the modelling of the effect of local superconducting correlations and long-range phase fluctuations on the spectral properties of high-temperature superconductors by reasoning that the PG is connected to the character of the excitations that are responsible for destroying superconductivity \cite{giob}.

Femtosecond spectroscopy is a tool for studying the temperature dependence of gaps, for example in Ref.\cite{demsar2} the temperature independence of the PG and dependence of the SG in Y$_{1-x}$CaBa$_2$Cu$_3$O$_{7-\delta}$ and HgBa$_2$Ca$_2$Cu$_3$O$_{8+\delta}$ was found. Raman spectroscopies have also been able to find two energy scales \cite{letacon}.
ARPES has provided valuable information about cuprates, ARPES performed on Bi2212 \cite{norman, tanaka, kanigel, leeet}, Bi2201 \cite{kondo}, LSCO \cite{yoshi}, LBCO \cite{valla} and CaNaCuOCl \cite{shenet} has verified the presence of two energy gaps in cuprates.

Scanning tunnelling microscopy (STM) offers a powerful technique to look at the doping, temperature and spatial dependence of the DOS with high resolution. It is sensitive to the DOS near the Fermi energy and to a gap in the quasiparticle excitation spectrum. 
  Extrinsic tunnelling experiments have left us with many questions regarding the properties of cuprates. STM tunnelling spectra exhibit an SG and PG \cite{langetal, howald, machetal} whose origin currently remains unaccounted for, despite many ideas partially discussed above. STM results on single crystals of Bi2212 (for example \cite{
mcelroy, gomesetal, panetal}) and LSCO \cite{katoetal, tkato} have demonstrated the temperature, doping and spatial dependence of the SG and PG. In particular, in NS tunnelling, Kato et al \cite{tkato} found that the PG is not uniform in real-space and its spatial average increases with decreasing hole concentration in spite of suppression of critical temperature. 
 On the other hand a smaller gap (presumable SG) is uniform across the sample and is less doping dependent.

\begin{figure}
\begin{center}
\includegraphics[angle=-90, width=0.5\textwidth]{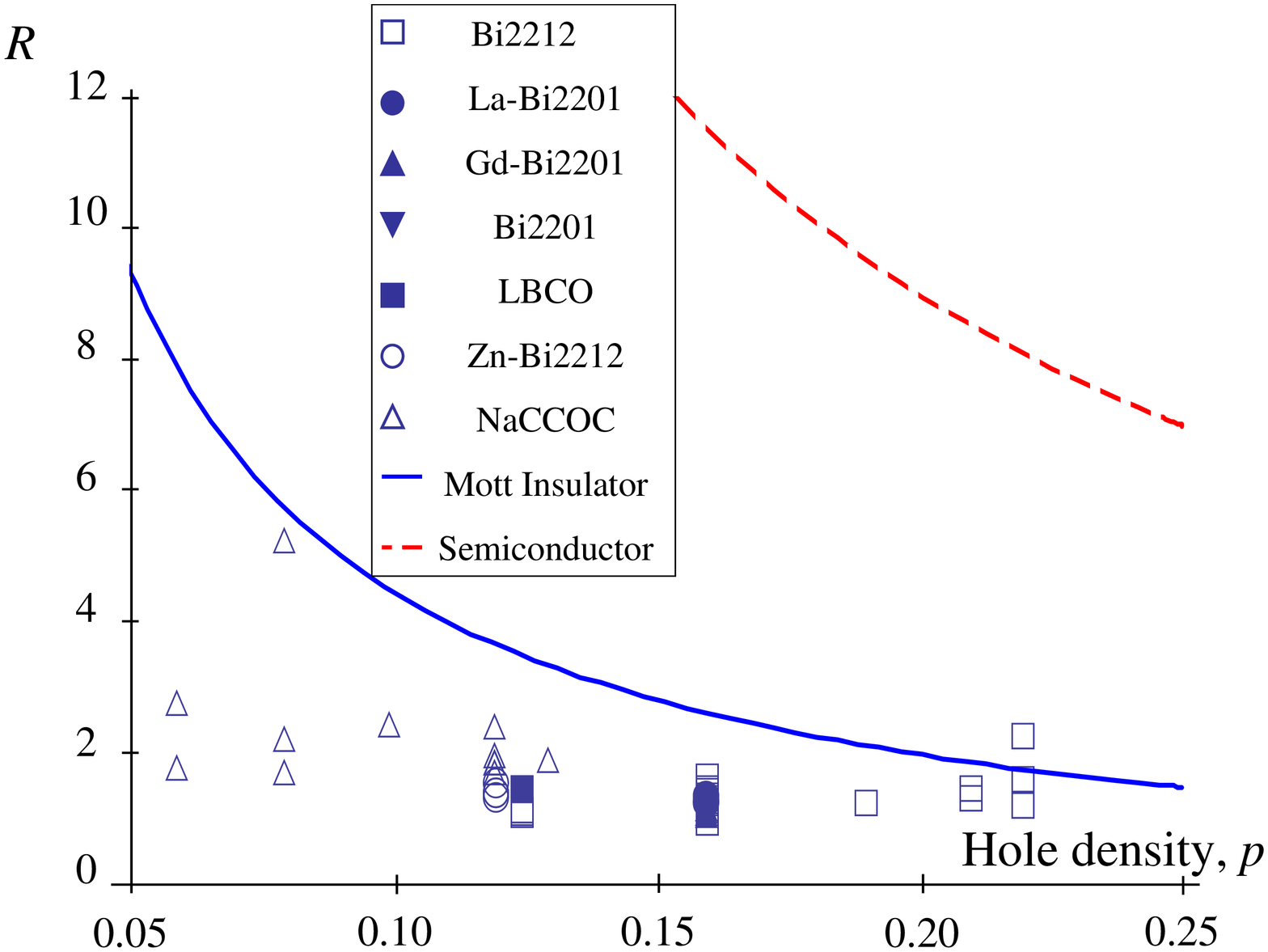}
 \caption{\small{(Colour online) Ratio of the negative bias NS tunnelling conductance to the positive bias $R=I_{NS}(-100)/I_{NS}(100)$, integrated from $0$ to $\mp100$meV respectively, carried out for some cuprate superconductors \cite{papersforgraph, parker2010} over a wide range of atomic hole density. The two curves express the asymmetry you would expected to see from a Mott insulator (solid blue) and conventional semiconductor (dashed red)  without electron hopping.}}\label{mottsemir}
\end{center}
\end{figure}

Intrinsic (superconductor-superconductor, SS) tunnelling experiments on small Bi2212 \cite{krasetal, suzwat, zhao, kras} and LSCO \cite{yurgens} mesas have found sharp quasiparticle peaks at the SG and broad humps representing the PG \cite{krasetal}. The PG exists above and below $T_c$ and can persist up to room temperature \cite{krasetal}. The advantages of intrinsic tunnelling are that it is a direct spectroscopic technique that avoids problems like surface deterioration \cite{zhao}, it probes the bulk electronic properties of samples, it offers high resolution whilst being mechanically stable so it is perfectly suited for temperature dependent studies of high-$T_c$ superconductors \cite{kras}. Break junction experiments also exhibit the SG \cite{miya, vedjan} in underdoped Bi2212 samples, the coexistence of the SG and the PG is seen in Bi2201 \cite{vedeneev} and Bi2212 \cite{xuantao} and quasiparticle energy gaps are found in Bi2212 tunnelling spectra \cite{miyakawa, vedmaud}.
\begin{figure}
\begin{center}
\includegraphics[width=0.25\textwidth]{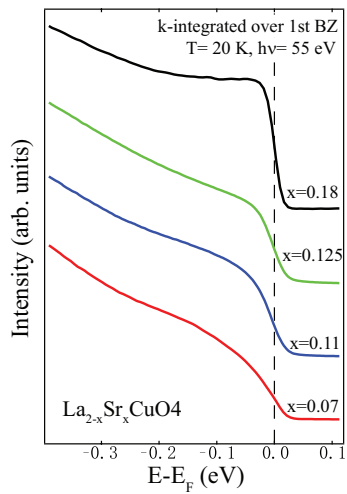}
 \caption{\small{(Colour online) Momentum integrated photoemission over the first Brillouin zone of La$_{2-x}$Sr$_x$CuO$_4$ \cite{RuihuaHe}, showing no signs of the van Hove singularity.}}\label{RHefig}
\end{center}
\end{figure}
One puzzling observation in cuprates is the significant asymmetry between the negative and positive bias conductance in NS tunnelling, which means the direction of the tunnelling carriers affects the tunnelling conductance. One possible explanation of the asymmetry was offered by the van Hove singularity of the DOS, a singularity in the DOS would give a visible hump in the ARPES data, however this does not seem possible since this singularity is not present in the momentum integrated photoemission \cite{RuihuaHe}, see Fig.\ref{RHefig} and Ref.\cite{rant}.
In the tunnelling spectra of conventional semiconductors or Mott-insulators asymmetry is expected. Consider a semiconductor where the number of electrons is twice the number of ions as each ion can accommodate a spin up and spin down electron. Removing $X$ electrons from the sample leaves the number of electrons as $2N-X$. When positive bias is applied to the sample, the electrons tunnel from the tip to the sample, application of negative bias gives tunnelling in the opposite direction. The probability of an electron tunnelling is, for negative bias (sample to tip), proportional to the number of electrons available this is $2N-X$, and for positive bias (tip to sample) the probability of tunnelling is proportional to the number of holes available in the sample for the electrons in the tip to tunnel to, this is $X$. The ratio of the integrated negative and positive conductance is given  by $R=(2-x)/x$, where $x=X/N$. Similarly we can consider a Mott insulator where the Coulomb repulsion is so strong that the number of electrons is equal to the number of ions as each ion can accommodate just one electron. Removing $X$ electrons from the sample leaves the number of electrons as $N-X$. Following the same idea as for the semiconductor, we have $R=(1-x)/2x$ \cite{rant}. This is a very basic formulation ignoring any electron  hopping, it is used to give us an idea of the magnitude of the asymmetry we can expect to see in a semiconductor or Mott insulator. It can be seen in Fig.\ref{mottsemir} that although both insulators exhibit asymmetry, they do not account for the magnitude of asymmetry seen in STM experiments with cuprate superconductors without the consideration of disorder and matrix elements \cite{ourPRL}.
Despite intensive research, a detailed microscopic theory capable of describing unusual ARPES and tunnelling data have remained elusive and so the relationship between the SG and PG has remained unknown. A detailed and consistent interpretation of SG and PG could shed light on the key  pairing interaction in cuprate superconductors.

High values of $T_c$ and small isotope effect on $T_c$ in optimally doped YBa$_2$Cu$_3$O$_{6.9}$ led some authors to conclude that the pairing interaction between electrons cannot be mediated by phonons. However experiments \cite{franck2, bournemannmor} showed that a partial substitution of Yttrium by Praseodymium, or of Barium by Lanthanum leads to the isotope effect simultaneously with the decrease of $T_c$: either these substituted compounds have a different mechanism of superconductivity or the mechanism is always phonons and the absence of the isotope effect in YBCO is due to something else.
In favour of the last option are the tunnelling spectra at higher voltages of NCCO \cite{tralshavala} and BSCCO \cite{shina, aaab}.
Evidence from the doping dependent oxygen effect (OIE) on $T_c$ and the substantial OIE on the carrier mass suggests a strong EPI in cuprate superconductors, where lattice vibrations play a significant but unconventional role in high temperature superconductivity, see Ref.\cite{alexzhao} and references therein.

In this paper we develop a theory of NS and SS tunnelling in the bosonic and cuprate superconductors extending our previous brief report \cite{ourPRL}.
The theory is based on the assumption that the EPI is strong enough in cuprates and similar ionic charge-transfer insulators to form small mobile bipolarons, which has been convincingly  supported by a  number of experimental observations \cite{asabluebook}. To clarify the terminology we first introduce bosonic superconductivity (Section \ref{bosonicsuperconductivity}). The results on the NS and SS tunnelling are presented in Sections \ref{NStunnelling} and \ref{SStunnelling}, respectively.

\section{Bosonic (bipolaronic) Superconductivity and Cuprate Band Structures}\label{bosonicsuperconductivity}
The BCS theory \cite{bcs} is capable of successfully describing the superconducting properties of elemental superconductors with a small EPI strength. The theory was modified in 1960 to give a strong coupling theory \cite{eliashberg} describing the properties of intermediate-coupling superconductors (the difference between a weak and strong-coupling superconductor is given by the electron-phonon coupling constant, $\lambda$ \cite{mcmillan}).
The mean-field BCS-Eliashberg theory is applied when the electron correlation length is large compared to the distance between them. The mechanism behind superconductivity is the momentum-space pairing of electrons through electron-phonon interactions. It was first realised by Fr\"ohlich in 1950 that electrons could be attracted to one another through their interactions with phonons; he suggested superconductivity was instigated by EPI.
The isotope effect observed experimentally verified Fr\"ohlich's proposition that EPI causes superconductivity.

When the coupling constant is increased above $\lambda\approx 1$, the kinetic energy of electrons becomes small compared with the potential energy from the local lattice deformation, thus all electrons in the Bloch band become dressed with phonons (for a recent review see
\cite{alexdev}). The electron becomes a quasiparticle, a small polaron, which can propagate through the lattice in a narrow (polaronic) band together with the lattice deformation.
For a further extension of the BCS theory towards a strong interaction between electrons and ion vibrations, $\lambda>1$, it was predicted that instead of Cooper pairs, a charged Bose gas of tightly bound small bipolarons would be evident \cite{alexran} with a polaronic BCS-like high-$T_c$ superconductivity in the crossover region \cite{Alexzhfiz}. These bipolarons are real-space pairs of two electrons with their phonon cloud.

Different from Cooper pairs in the momentum-space, the ground-state of the strongly coupled electrons and phonons is the real-space pairing of these single polarons into bosonic bipolarons where they form a condensate which can be described as a charged Bose-liquid on a lattice if the carrier density is small enough to avoid their overlap \cite{asabluebook}. In the superconducting state, if the temperature is finite, not all the polarons will condense and those that have not condensed interact with the condensate through the same potential that binds them together. The single-particle Hamiltonian is described as \cite{alexandreev}:
\begin{equation}
H_0=\sum_{\nu}\left[\xi_{\nu}p_{\nu}^{\dagger}p_{\nu}+\frac{\Delta_{c\nu}}{2}\left(p_{\bar{\nu}}^{\dagger}p_{\nu}^{\dagger}+\text{h.c.}\right)\right],
\end{equation}
where $\xi_{\nu}=E_{\nu}-\mu$, $E_{\nu}$ is the normal-state single polaron energy spectrum in the crystal field and disorder potentials renormalized by EPI and spin fluctuations and $\Delta_{c\nu}=-\Delta_{c\bar{\nu}}$ is the coherent potential proportional to the square root of the condensate density, $\Delta_c\propto\sqrt{n_c(T)}$. The operators $p_{\nu}^{\dagger}$ and $p_{\bar{\nu}}^{\dagger}$ create a polaron in the single particle quantum state $\nu$ and in the time reversed state $\bar{\nu}$ respectively.

As in the BCS case, the single-particle energy spectrum $\epsilon_{\nu}$ is found by applying the Bogoliubov transformation to diagonalise the Hamiltonian, which is thus written:
\begin{equation}
H_0=\sum_{\nu}\epsilon_{\nu}\left(\alpha_{\nu}^{\dagger}\alpha_{\nu}+\beta_{\nu}^{\dagger}\beta_{\nu}\right),
\end{equation}
where $p_{\nu}=u_{\nu}\alpha_{\nu}+v_{\nu}\beta_{\nu}^{\dagger}$, $p_{\bar{\nu}}=u_{\nu}\beta_{\nu}-v_{\nu}\alpha_{\nu}^{\dagger}$, $\epsilon_{\nu}=\sqrt{\xi_{\nu}^2+\Delta_{c\nu}^2}$ with $u_{\nu}^2,\ v_{\nu}^2=\frac{1}{2}\left(1\pm\frac{\epsilon_{\nu}}{\xi_{\nu}}\right)$. This spectrum is different to the BCS quasiparticles because the chemical potential ($\mu$), is negative with respect to the bottom of the single-particle band, $\mu=-\Delta_p$. A single-particle gap $\Delta$, is defined as the minimum of the single-quasiparticle energy spectrum. Without disorder, for a point-like pairing potential with the $s$-wave coherent gap, $\Delta_{c\boldsymbol{k}}\approx\Delta_c$, one has \cite{alexandreev}:
\begin{equation}
\Delta(T)=\sqrt{\Delta_p^2+\Delta_c(T)^2}.
\end{equation}
The full gap varies with temperature from $\Delta(0)=\sqrt{\Delta_p^2+\Delta_c(0)^2}$ at zero temperature to the temperature independent $\Delta=\Delta_p$ above $T_c$, which qualitatively describes some earlier and more recent \cite{kras} observations including Andreev reflection in cuprates (\cite{alexandreev} and references therein).

\begin{figure}
\begin{center}
\includegraphics[width=0.5\textwidth]{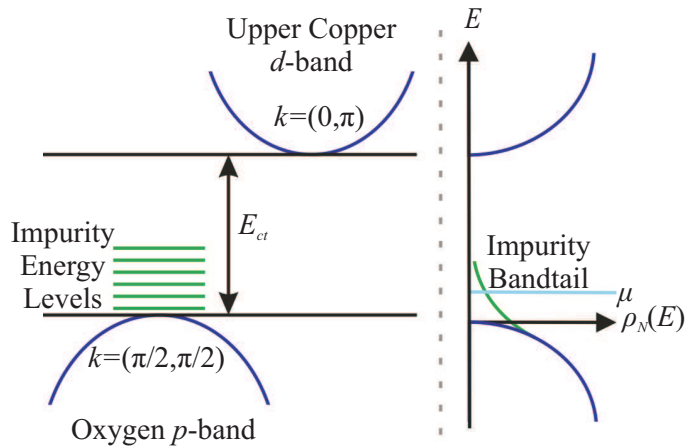}
 \caption{\small{(Colour online) The``LDA+GTB" band structure is shown on the left with the impurity levels indicated, these impurity levels cause the bandtail in the cuprate superconductor DOS on the right.}}\label{dosen}
\end{center}
\end{figure}

In this paper we adopt the parent band structure based upon a Mott or any semiconductor insulator structure. The copper band is split into two; an upper and lower band which makes the Mott insulator, the gap between the two is approximately $5-8eV$. The semiconductor oxygen $p$-band lies within this gap and its charge transfer gap is approximately $1-2eV$, so we consider a parent lattice that is half Mott-insulator, half semiconductor. We have assumed the local density approximation and generalised tight binding (``LDA+GTB") band structure \cite{korsgavetal}, see Fig.\ref{dosen}.

Here, following Ref.\cite{alexkim} we amend this structure with the impurity bandtails. When the cuprate is doped an impurity ion locally introduces a distinct energy level within the charge transfer gap. The random spatial distribution of impurities when doping causes a bandtail effect of the DOS, similar to that of a heavily doped semiconductor. This band structure explains the charge transfer gap, sharp quasiparticle peaks near $\left(\frac{\pi}{2},\frac{\pi}{2}\right)$ of the Brillouin zone and a high energy waterfall observed by ARPES in underdoped cuprate superconductors \cite{alexkim}. Only the impurity states with binding energy below $\mu=0$ contribute at $T=0$K. It has been suggested that the band structure should be metallic due to Fermi arcs seen by ARPES which could form part of a large Fermi surface; however this does not take into account strong correlations and the existence of the charge-transfer gap over a wide range of dopings \cite{korsgavetal, mardasetal}.
For tunnelling to be possible, a state must be occupied by a carrier and at the same energy level on the other material a state must be vacant. The probability of a single-particle state being occupied is given by the Fermi-Dirac distribution. We consider single-particle tunnelling only, this is in key with STM results as they measure one particle tunnelling only (rather than Josephson tunnelling where the tunnelling of pairs can occur). The tunnelling process is described by the standard perturbation theory, where the tunnelling Hamiltonians are perturbations, then the Fermi-Dirac golden rule (FDGR) is applied.
%
%
\section{NS Tunnelling}\label{NStunnelling}
\begin{figure}
\begin{center}
\includegraphics[width=0.6\textwidth]{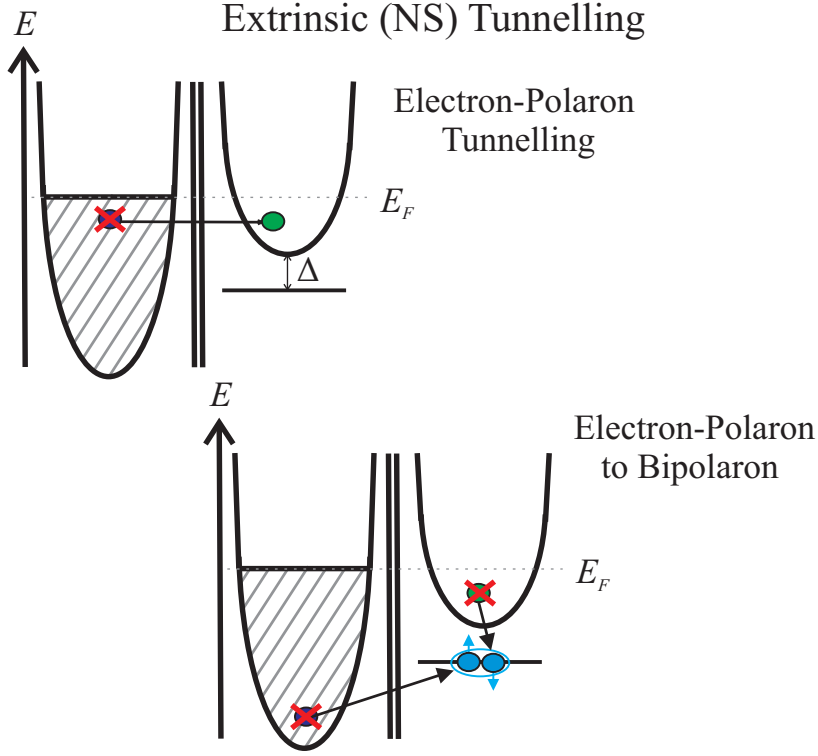}
 \caption{\small{(Colour online) Cartoon demonstrating the two possible single-particle tunnelling scenarios. The first is the annihilation of an electron in the superconductor on the left and the creation of a polaron on the right, as described in the first part of the tunnelling Hamiltonian. The second illustrates the tunnelling process with the involvement of a bipolaron. For normal-metal to superconductor tunnelling, this is the annihilation of an electron in the metal and the annihilation of a polaron in the superconductor with the creation of a composed boson, this is described in the second term of Eq.(\ref{nsham}). Energy is conserved in the tunnelling process.} }\label{exttun}
\end{center}
\end{figure}
To find the NS tunnelling Hamiltonian each different tunnelling scenario needs to be considered, Fig.\ref{exttun}.  Suppose on the left we have the metallic tip with undressed carriers as opposed to the polarons and bipolarons on the right, superconducting side. This means for tunnelling left to right we have the annihilation of a free carrier on the left accompanied by the creation of a polaron on the right. Alternatively we might have the annihilation of a carrier on the left and the annihilation of a polaron on the right with the creation of a bipolaron on the right. This can be expressed in terms of the Hamiltonian \cite{alexandrov}: (``h.c." is the Hermitian conjugate, describing the tunnelling in opposing direction)
\begin{equation}
H_{NS}=P\sum_{\nu\nu^{\prime}}p_{\nu^{\prime}}^{\dagger}c_{\nu}+\frac{B}{\sqrt{N}}\sum_{\nu\nu^{\prime}\eta^{\prime}}b_{\eta^{\prime}}^{\dagger}p_{\bar{\nu}^{\prime}}c_{\nu}+\text{h.c..}\label{nsham}
\end{equation}
Here $c_{\nu}$ and $b_{\eta^{\prime}}^{\dagger}$ are the annihilation of a carrier in the metallic tip in state $\nu$ and the creation of a composed boson in the superconductor in state $\eta^{\prime}$ respectively, $N$ is the number of lattice cells. $P$ and $B$ are tunnelling matrix elements respectively with and without the involvement of a bipolaron. Generally $B\gtrsim P$, because the presence of an additional hole lowers the tunnelling barrier for an injection of the electron \cite{alexandrov}. Using the Bogoliubov coefficients
\begin{equation}
u_{\nu}^2=\frac{1}{2}\left(1+\frac{\xi_{\nu}}{\epsilon_{\nu}}\right);\ v_{\nu}^2=\frac{1}{2}\left(1-\frac{\xi_{\nu}}{\epsilon_{\nu}}\right),\label{boggers}
\end{equation}
to replace the polaron operators with linear combinations of the quasiparticle operators yields:
\begin{equation}
H_{NS}=P\sum_{\nu\nu^{\prime}}\left(u_{\nu^{\prime}}\alpha_{\nu^{\prime}}^{\dagger}+v_{\nu^{\prime}}\beta_{\nu^{\prime}}\right)c_{\nu}+\frac{B}{\sqrt{N}}\sum_{\nu\nu^{\prime}\eta^{\prime}}b_{\eta^{\prime}}^{\dagger}\left(u_{\nu^{\prime}}\beta_{\nu^{\prime}}-v_{\nu^{\prime}}\alpha_{\nu^{\prime}}^{\dagger}\right)c_{\nu}+\text{h.c..}
\end{equation}
Using the Fermi-Dirac Golden Rule
\begin{equation}
W=\frac{2\pi}{\hbar}\left|\left\langle \left\{n\right\}\left|H_{\text{tun}}\right|\left\{0\right\}\right\rangle\right|^2\delta(E_n-E_0),
\end{equation}
where the initial state is $0$ and the final is $n$, yields:
\begin{eqnarray}
W_{NS}^{\text{in}}&=&\frac{2\pi P^2}{\hbar}\sum_{\nu\nu^{\prime}}\left[u_{\nu^{\prime}}^2(1-f_{\nu^{\prime}})F_{\nu}\delta(\xi_{\nu}+eV-\epsilon_{\nu^{\prime}})+v_{\nu^{\prime}}^2f_{\nu^{\prime}}F_{\nu}\delta(\xi_{\nu}+eV+\epsilon_{\nu^{\prime}})\right]\cr
& &+\frac{2\pi B^2}{N}\sum_{\nu\nu^{\prime}\eta^{\prime}}(1+n_{\eta^{\prime}})\left[u_{\nu^{\prime}}^2f_{\nu^{\prime}}F_{\nu}\delta(E_{\eta^{\prime}}-\xi_{\nu}-eV-\epsilon_{\nu^{\prime}})+v_{\nu^{\prime}}^2(1-f_{\nu^{\prime}})F_{\nu}\delta(E_{\eta^{\prime}}-\xi_{\nu}-eV+\epsilon_{\nu^{\prime}})\right],
\end{eqnarray}
\begin{eqnarray}
W_{NS}^{\text{out}}&=&\frac{2\pi P^2}{\hbar}\sum_{\nu\nu^{\prime}}\left[u_{\nu^{\prime}}^2f_{\nu^{\prime}}(1-F_{\nu})\delta(\xi_{\nu}+eV-\epsilon_{\nu^{\prime}})+v_{\nu^{\prime}}^2(1-f_{\nu^{\prime}})(1-F_{\nu})\delta(\xi_{\nu}+eV+\epsilon_{\nu^{\prime}})\right]\cr
& &+\frac{2\pi B^2}{N}\sum_{\nu\nu^{\prime}\eta^{\prime}}n_{\eta^{\prime}}\left[u_{\nu^{\prime}}^2(1-f_{\nu^{\prime}})(1-F_{\nu})\delta(E_{\eta^{\prime}}-\xi_{\nu}-eV-\epsilon_{\nu^{\prime}})+v_{\nu^{\prime}}^2f_{\nu^{\prime}}(1-F_{\nu})\delta(E_{\eta^{\prime}}-\xi_{\nu}-eV+\epsilon_{\nu^{\prime}})\right].
\end{eqnarray}
Here $W^{\text{in}}$ and $W^{\text{out}}$ are transition rates in and out of the superconductor, $f_{\nu^{\prime}}=1/(e^{\epsilon_{\nu^{\prime}}/K_BT}+1)$ is the single quasiparticle distribution function, $n_{\eta^{\prime}}$ is the bipolaron (Bose) distribution function, $F_{\nu}=1/(e^{\xi_{\nu}/K_BT}+1)$ describes the distribution of carriers in the normal metal, $V$ is the voltage drop across the junction and the bipolaron chemical potential in the left superconductor differs from the right one by $2eV$. To find the current we use the equation
\begin{equation}
I=e(W_{\text{in}}-W_{\text{out}}),\label{current}
\end{equation}
which gives \cite{ourPRL}:
\begin{eqnarray}
I_{NS}(V)&=&\frac{2\pi e P^2}{\hbar}\sum_{\nu\nu^{\prime}}\left[u_{\nu^{\prime}}^2\left(F_{\nu}-f_{\nu^{\prime}}\right)\delta(\xi_\nu+eV-\epsilon_{\nu^{\prime}})+v_{\nu^{\prime}}^2 \left(F_{\nu}+f_{\nu^{\prime}}-1\right)\delta (\xi_\nu+eV+\epsilon_{\nu^{\prime}})\right]\cr
& & + \frac{2\pi e B^2}{\hbar}\sum_{\nu \nu^{\prime}}\left\{u_{\nu^{\prime}}^2\left[F_{\nu}f_{\nu'}-(x/2) (1-F_\nu-f_{\nu^{\prime}})\right] \delta
(\xi_\nu+eV+\epsilon_{\nu^{\prime}})\right.\cr
& &\ \ \ \ \ \ \ \ \ \ + \left.v_{\nu'}^2\left[F_{\nu}(1-f_{\nu^{\prime}})+(x/2) (F_\nu-f_{\nu^{\prime}})\right]\delta(\xi_\nu+eV-\epsilon_{\nu^{\prime}})\right\},
\end{eqnarray}
where $x/2$ is the atomic density of composed bosons in the superconductor. The boson energy dispersion is neglected here for more transparency, assuming that they are sufficiently heavy for their bandwidth to be relatively small. Using $\sum_{\nu}\rightarrow\int_{-\infty}^{\infty}\rho_M(\xi)d\xi$, $\sum_{\nu^{\prime}}\rightarrow\int_{-\infty}^{\infty}\rho_N(\xi^{\prime})d\xi^{\prime}$, and neglecting the energy dependence of the metallic DOS $\rho_M(\xi)$ since near the Fermi energy it is approximately a constant, we obtain:
\begin{eqnarray}
I_{NS}&&=\frac{2\pi e P^2 \rho_M}{\hbar}\int d\xi\int d\xi^{\prime}\rho_N(\xi^{\prime})
\left\{u^2(\xi^{\prime})\left[F(\xi-eV)-f(\epsilon^{\prime})\right]\delta(\xi-\epsilon^{\prime})
+v^2(\xi^{\prime})\left[F(\xi-eV)+f(\epsilon^{\prime})-1\right]\delta(\xi+\epsilon^{\prime})\right\}\cr
&&+\frac{2\pi e B^2 \rho_M}{\hbar}\int d\xi\int d\xi^{\prime} \rho_N(\xi^{\prime})
\left\{u^2(\xi^{\prime})\left[F(\xi-eV)f(\epsilon^{\prime})-\frac{x}{2}\left(1-F(\xi-eV)-f(\epsilon^{\prime})\right)\right]
\delta\left(\xi+\epsilon^{\prime}\right)\right.\cr
&&\ \ \ \ \ \ \ \ \ \ \ \ \ \ \ \ \ \ \ \ \ \ \ \ +\left.v^2(\xi^{\prime})\left[F(\xi-eV)\left(1-f(\epsilon^{\prime})\right)+\frac{x}{2}\left(F(\xi-eV)-f(\epsilon^{\prime})\right)\right]\delta(\xi+\epsilon^{\prime})\right\},
\end{eqnarray}
where $\rho_N(\xi)$ is  the normal-state single-particle DOS in the doped charge-transfer insulator. At zero temperature, the Fermi-Dirac distribution becomes a step function, $F(\xi-eV)\rightarrow\Theta(eV-\xi)$ and $f(\epsilon^{\prime})=0$. To find the conductance, the current is differentiated with respect to the voltage and we have:
\begin{eqnarray}
\sigma_{NS}=&&\frac{2\pi e P^2 \rho_M}{\hbar}\int d\xi\int d\xi^{\prime} \rho_N(\xi^{\prime})
\left[u^2(\xi^{\prime})\delta(eV-\xi)\delta(\xi-\epsilon^{\prime})+v^2(\xi^{\prime})\delta(eV-\xi)\delta(\xi+\epsilon^{\prime})\right]\cr
&&+\frac{2\pi e B^2 \rho_M}{\hbar}\int d\xi\int d\xi^{\prime} \rho_N(\xi^{\prime})\left[u^2(\xi^{\prime})\delta(eV-\xi)\delta(\xi+\epsilon^{\prime})\frac{x}{2}\right.\cr
&&\ \ \ \ \ \ \ \ \ \ \ \ \ \ \ \ \ \ \ \ \ \ \ \
+\left.v^2(\xi^{\prime})\delta(eV-\xi)\delta(\xi-\epsilon^{\prime})\left(1+\frac{x}{2}\right)\right].
\end{eqnarray}
Using the Bogoliubov coefficients defined earlier, Eq.(\ref{boggers}), we find:
\begin{eqnarray}
\sigma_{NS}&\propto&\Theta(eV-\Delta_c)\left\{\rho_S(eV)\left(2+\frac{x}{2}\right)\left[\rho_N(\sqrt{(eV)^2-\Delta_c^2})+\rho_N(-\sqrt{(eV)^2-\Delta_c^2})\right]\right.\cr
& &\left.\ \ \ \ \ \ \ \ \ \ \ \ \ \ \ \ \ \ \ \ \ \ \ \ \ \ \ \ \ \ \ \ \ \ \ +\frac{x}{2}\left[\rho_N(-\sqrt{(eV)^2-\Delta_c^2})-\rho_N(\sqrt{(eV)^2-\Delta_c^2})\right]\right\}\cr
& &+\Theta(-eV-\Delta_c)\left\{\rho_S(eV)\left(1+\frac{x}{2}\right)\left[\rho_N(\sqrt{(eV)^2-\Delta_c^2})+\rho_N(-\sqrt{(eV)^2-\Delta_c^2})\right]\right.\cr
& &\left.\ \ \ \ \ \ \ \ \ \ \ \ \ \ \ \ \ \ \ \ \ \ \ \ \ \ \ \ \ \ \ \  +\left(1-\frac{x}{2}\right)\left[\rho_N(-\sqrt{(eV)^2-\Delta_c^2})-\rho_N(\sqrt{(eV)^2-\Delta_c^2})\right]\right\}
\end{eqnarray}
where $\Theta(x)$ is the Heaviside step-function, $\rho_S(E)=E/\sqrt{E^2-\Delta_c^2}$ for $s$-wave symmetry of the coherent gap which does not depend on the quantum number $\nu$ and we assume here $P=B$ for more transparency.
%
%
Similarly, the conductance can be found for $d$-wave symmetry, where the coherent gap is now a function of $\nu$, $\Delta_{c\nu}=\Delta_0\text{cos}2\phi$, $\phi$ is an angle along a constant energy contour. The DOS in the superconducting state is given by \cite{wonandmaki}:
\begin{equation}
\rho_S(E)=\frac{2}{\pi}\left[\Theta\left(1-\frac{E}{\Delta_0}\right)\frac{E}{\Delta_0}K\left(\frac{E}{\Delta_0}\right)+\Theta\left(\frac{E}{\Delta_0}-1\right)K\left(\frac{\Delta_0}{E}\right)\right].
\end{equation}
Then, if the tail-width ($\Gamma$) is large compared with the coherent gap amplitude, $\Gamma > \Delta_0$, Eq.(13) yields
\begin{equation}
\sigma_{NS}\propto A^+\rho_S(\left|eV\right|)\left[\rho_N(-eV)+\rho_N(eV)\right]+A^-\left[1-\frac{2}{\pi}\text{arccos}\left(\frac{\left|eV\right|}{\Delta_0}\right)\Theta\left(1-\frac{\left|eV\right|}{\Delta_0}\right)\right]\left[\rho_N(-eV)-\rho_N(eV)\right],\label{dwavensconductance}
\end{equation}
where $A^{\pm}=1\pm B^2\left(\Theta(-eV)+\frac{x}{2}\right)/P^2$.
%
%
\begin{figure}
\begin{center}
\includegraphics[width=0.5\textwidth]{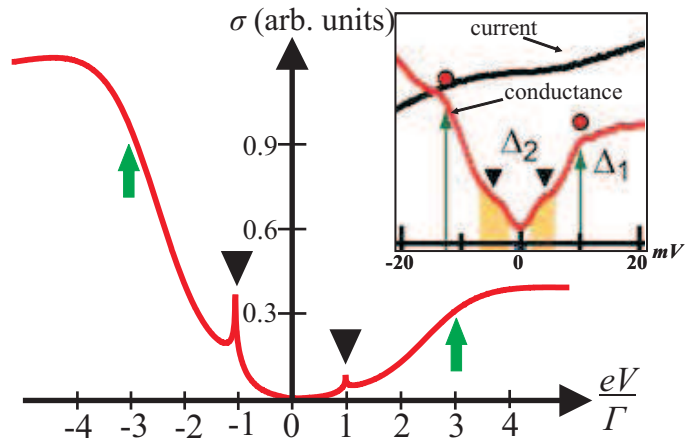}
 \caption{\small{(Colour online) Theoretical extrinsic conductance, Eq.(\ref{dwavensconductance}) for $\Delta_0=\Gamma$, $\Delta_p=2.7\Gamma$ and $B=2.65P$. The SG and PG are indicated by the triangles and arrows respectively. Inset shows a representative STS spectrum of La$_{2-x}$Sr$_x$CuO$_4$ at $4.2$K \cite{tkato}.}}\label{theonscond}
\end{center}
\end{figure}
As aforementioned, the doping of impurities in the cuprates causes a bandtailing effect in the normal-state DOS. We have used a model DOS to reflect the shape of our DOS which is given by:
\begin{equation}
\frac{\rho_N(E)}{\rho_b}=\frac{1}{2}\left[1+\tanh\left(\frac{E-\Delta_p}{\Gamma}\right)\right].\label{modelDOS}
\end{equation}
This model $\rho_N(E)$ reflects the characteristic energy dependence of the DOS in disordered doped insulators, which is a constant $\rho_b$ above the two-dimensional band edge, and an exponent deep in the tail (see Fig.\ref{dosen}).

No matter what symmetry the superconducting gap is, the above equations capture all the unusual signatures of the extrinsic experimental tunnelling conductance in underdoped cuprates, such as the low energy coherent SG, the high energy PG and the asymmetry, see Fig.\ref{theonscond}. In the case of atomically resolved STS one should replace the averaged $\rho_N(E)$ in the above equations with a local bandtail DOS, $\rho_N(E,\boldsymbol{r})$, which is dependent on the position of the tip on the scanned area due to a nonuniform dopant distribution. As a result, the PG shows nanoscale inhomogeneity, while the low energy SG is spatially uniform as observed \cite{tkato}. Increasing doping level tends to diminish the bipolaron binding energy, $\Delta_p$, since the pairing potential becomes weaker due to a partial screening of EPI with low frequency phonons \cite{alekabmot}. However, the coherent gap $\Delta_c$, which is a product of the pairing potential and the squareroot of the carrier density \cite{alexandreev}, can remain about a constant or even increase with doping, as observed \cite{tkato, ourPRL}.
%
%
\section{SS Tunnelling}\label{SStunnelling}
 For SS tunnelling there are different issues to address, in particular, tunnelling in mesas has indicated a nonzero conductance at zero voltage near and above $T_c$, also a negative excess resistance has been observed, along with the PG and SG. All of these features are accounted for in our theory.
 \begin{figure}
\begin{center}
\includegraphics[width=0.45\textwidth]{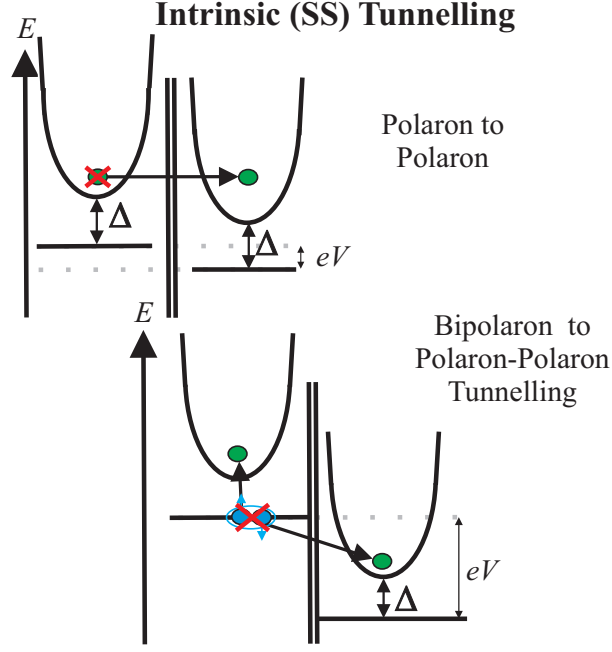}
 \caption{\small{(Colour online) Cartoon demonstrating  different single-particle tunnelling processes, the first is the annihilation of an electron in the superconductor on the left and the creation of a polaron on the right, the same as for NS tunnelling. This is described in the first part of the tunnelling Hamiltonian Eq.(\ref{ssham}). The second illustrates a tunnelling process involving a bipolaron, where on the left a bipolaron is annihilated into two polarons, one of these moves into the polaron band on the left, the other tunnels to the superconductor on the right, as described by the second term of Eq.(\ref{ssham}).}}
\end{center}
\end{figure}
The Hamiltonian \cite{alexandrov}:
\begin{equation}
H_{SS}=P\sum_{\nu\nu^{\prime}}p_{\nu^{\prime}}^{\dagger}p_{\nu}+\frac{B}{\sqrt{N}}\sum_{\nu\nu^{\prime}}\left(\sum_{\eta}p_{\nu^{\prime}}^{\dagger}p_{\bar{\nu}}^{\dagger}b_{\eta}+\sum_{\eta^{\prime}}b_{\eta^{\prime}}^{\dagger}p_{\nu}p_{\bar{\nu}^{\prime}}\right)+\text{h.c.},\label{ssham}
\end{equation}
describes the tunnelling of a single polaron from one bosonic superconductor to the other. The first term has no involvement of bipolarons and describes the annihilation of a polaron in state $\nu$ and creation of one in state $\nu^{\prime}$. The second term involves the decay of a composed boson into  two polarons, one remains in the same superconductor as the boson and is in state $\bar{\nu}$ (this is the time reversed state of $\nu$). The other tunnels into state $\nu^{\prime}$ in the other superconductor. The third term is the opposite to this with the annihilation of two polarons, one from each superconductor, that then combine to form a bipolaron. Again, only single-particle tunnelling is considered. Following the same procedure as for NS tunnelling (applying the FDGR and Eq.(\ref{current})) yields:
\begin{eqnarray}
I_{SS}(V)&=&\frac{2\pi e P^2}{\hbar}\sum_{\nu \nu^{\prime}}
\left[(u_{\nu}^2u_{\nu^{\prime}}^2+v_{\nu}^2 v_{\nu^{\prime}}^2) (f_{\nu}-f_{\nu^{\prime}})\delta(\epsilon_\nu+eV-\epsilon_{\nu^{\prime}})
+u_{\nu}^2v_{\nu^{\prime}}^2 (f_{\nu}+f_{\nu^{\prime}}-1)\right]
\cr
& &\ \ \ \ \ \ \ \ \ \ \ \ \ \ \ \ \ \ \ \ \ \ \ \ \ \ \ \ \ \ \ \ \ \ \ \ \ \ \ \ \ \ \ \ \ \ \ \ \ \ \ \ \ \ \ \ \ \ \ \ \ \ \ \ \ \ \ \ \ \times\left[\delta(\epsilon_\nu+eV+\epsilon_{\nu^{\prime}})-\delta(\epsilon_\nu-eV+\epsilon_{\nu^{\prime}})\right]\cr
& &+\frac{2\pi e B^2}{\hbar}\sum_{\nu \nu^{\prime}}\left[u_{\nu}^2u_{\nu^{\prime}}^2\left(\left(1-f_\nu-f_{\nu^{\prime}}\right)\frac{x}{2}-f_{\nu}f_{\nu^{\prime}}\right)+ v_{\nu}^2v_{\nu^{\prime}}^2
\left(\left(1-f_\nu-f_{\nu^{\prime}}\right)\frac{x}{2}+(1-f_{\nu})(1-f_{\nu^{\prime}})\right)\right]\cr
& &\ \ \ \ \ \ \ \ \ \ \ \ \ \ \ \ \ \ \ \ \ \ \ \ \ \ \ \ \ \ \ \ \ \ \ \ \ \ \ \ \ \ \ \ \ \ \ \ \ \ \ \ \ \ \ \ \ \ \ \ \ \ \ \ \ \ \ \ \ \times \left[\delta(\epsilon_\nu-eV+\epsilon_{\nu^{\prime}})-\delta(\epsilon_{\nu}+eV+\epsilon_{\nu^{\prime}})\right]\cr
& & \ \ \ \ \ \ \ \ \ \ \ \ \ \ \ \ +2u_{\nu}^2v_{\nu^{\prime}}^2 \left[(f_{\nu^{\prime}}-f_{\nu})\frac{x}{2}-f_{\nu}(1-f_{\nu^{\prime}})\right] \left[\delta(\epsilon_{\nu}+eV-\epsilon_{\nu^{\prime}})-\delta(\epsilon_\nu-eV-\epsilon_{\nu'})\right].\label{equationss1}
\end{eqnarray}
The boson energy dispersion is again dropped here, assuming that bipolarons are sufficiently heavy for their bandwidth to be relatively small.
%

\begin{figure}
\begin{center}
\includegraphics[angle=-90, width=0.6\textwidth]{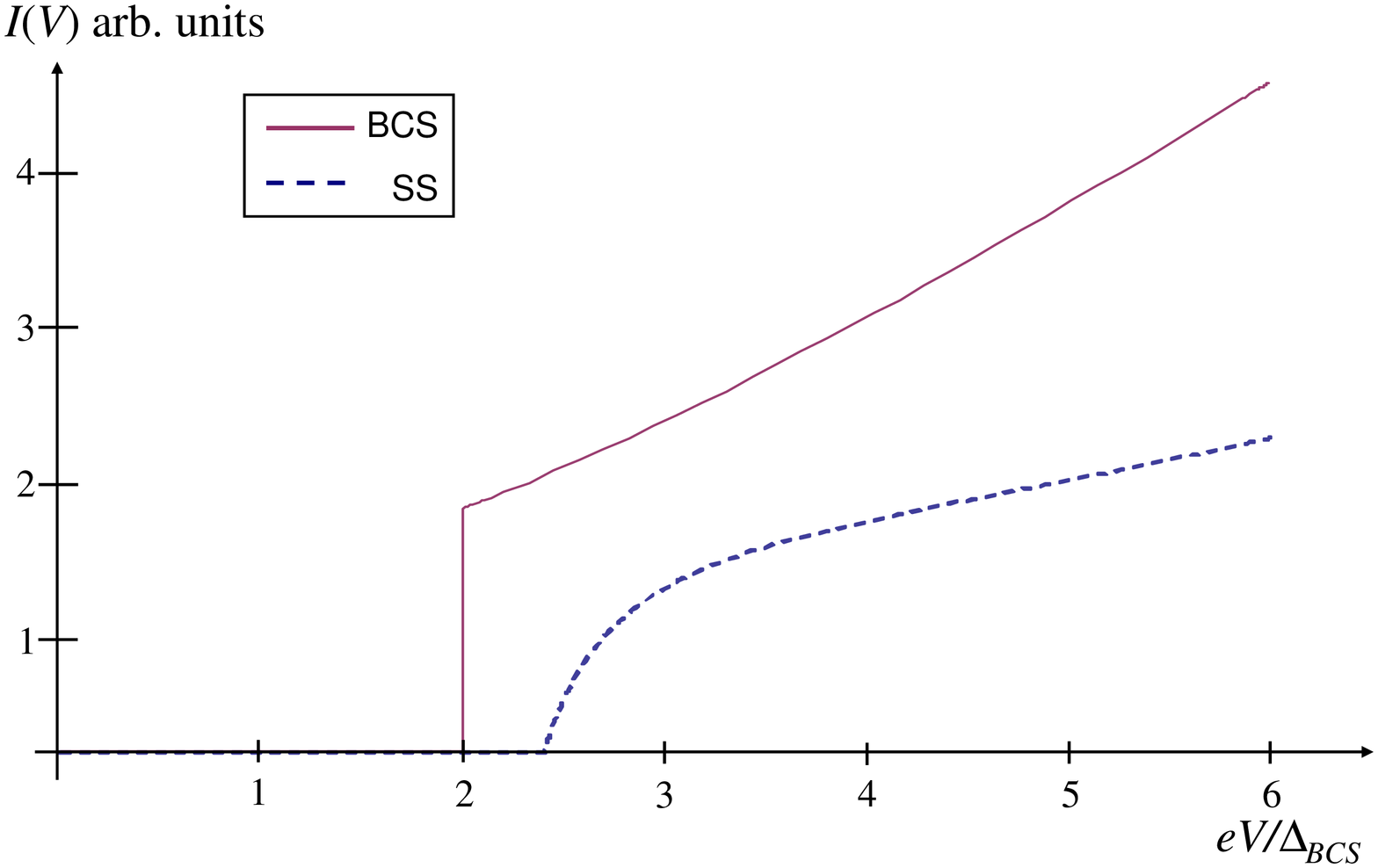}
 \caption{\small{(Colour online) Our theory for SS tunnelling in a `clean' bosonic superconductor, with no bandtail, compared to superconductor-superconductor tunnelling in the BCS theory \cite{mahan} at zero temperature for $P=B$, $x/2=0.16$, $\Delta_c=\Delta_{BCS}$ and $\Delta=1.2\Delta_c$.}}\label{mahantheory}
\end{center}
\end{figure}

First consider a clean bosonic superconductor, where there are no bandtails ($\Gamma=0$), compare its  $I(V)$ with the SS $I(V)$  in the BCS case for zero temperature when $f_{\nu}=0$. Refer back to Eq.(\ref{modelDOS}), the normal state DOS $\rho_N(\xi)$ becomes a step function as $\Gamma$, which is the width of the bandtail, tends to zero, $\rho_N(\xi)\rightarrow\Theta(\xi-\Delta_p);\ \rho_N(\xi^{\prime})\rightarrow\Theta(\xi^{\prime}-\Delta_p)$. Thus, integrating with respect to $\xi$ and $\xi^{\prime}$ we have for the current:
\begin{equation}
I_{SS}\propto2\Theta(eV-2\Delta)\left\{
\left(\frac{x}{2}+1\right)I_0+\frac{x}{2}(eV-2\Delta)-\left[\sqrt{(eV-\Delta)^2-\Delta_c^2}-\sqrt{\Delta^2-\Delta_c^2}\right]\right\},
\end{equation}
such that
\begin{equation}
I_0=\frac{(eV)^2}{eV+2\Delta_c}F(\arcsin(1-\beta),\alpha)-(eV+2\Delta_c)\left[F(\arcsin(1-\beta),\alpha)-E(\arcsin(1-\beta),\alpha)\right],
\end{equation}
where $\beta=\frac{2(\Delta-\Delta_c)}{eV-2\Delta_c}$, $\alpha=\frac{eV-2\Delta_c}{eV+2\Delta_c}$, we take $P=B$, $F(x,y)$ and $E(x,y)$ are incomplete elliptic integrals of the first and second kind respectively.
This is plotted in Fig.\ref{mahantheory} against what is expected for SS tunnelling in the BCS theory \cite{mahan}:
\begin{equation}
I_{SS}\propto\Theta(eV-2\Delta_{\text{BCS}})\left\{\frac{(eV)^2}{eV+2\Delta_{\text{BCS}}}K(\alpha)-(eV+2\Delta_{\text{BCS}})\left[K(\alpha)-E(\alpha)\right]\right\},
\end{equation}
where $\alpha=\frac{eV-2\Delta_{\text{BCS}}}{eV+2\Delta_{\text{BCS}}}$, $K(\alpha),\ E(\alpha)$ are complete elliptic integrals (see \cite{mahan} for more details.)
%

\begin{figure}
\begin{center}
\includegraphics[angle=-90, width=0.45\textwidth]{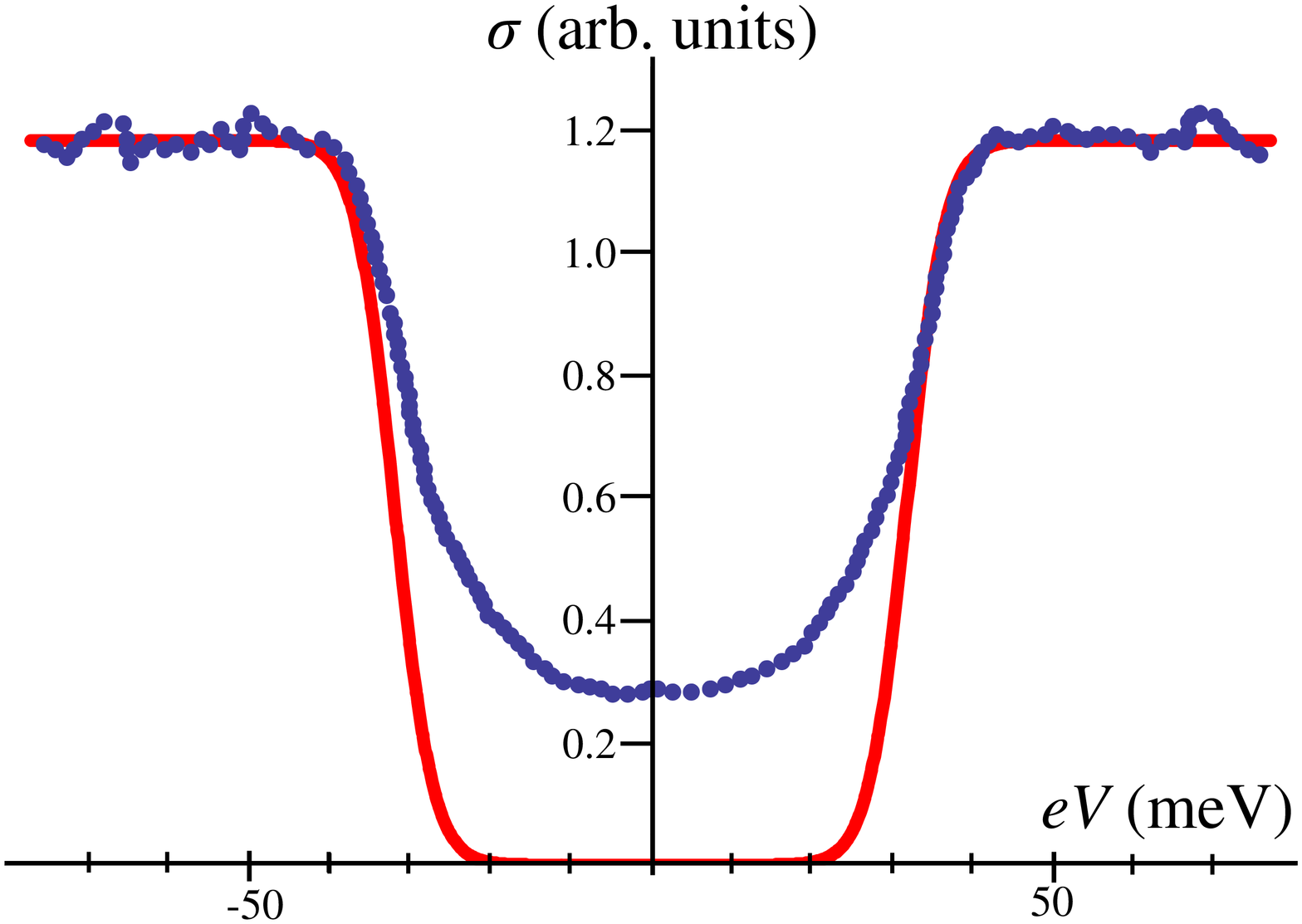}
 \caption{\small{(Colour online) Approximate normal-state tunnelling conductance of bosonic superconductor (solid line (red)) with $\Gamma=3.2$meV and $\Delta_p=16$meV \cite{ourPRL} compared with experimental conductance \cite{kras} in mesas of Bi2212 ($T_c=95$K at $T=85$K).}}\label{normalstatessconductance}
\end{center}
\end{figure}

\begin{figure}
\begin{center}
\includegraphics[angle=-90, width=0.6\textwidth]{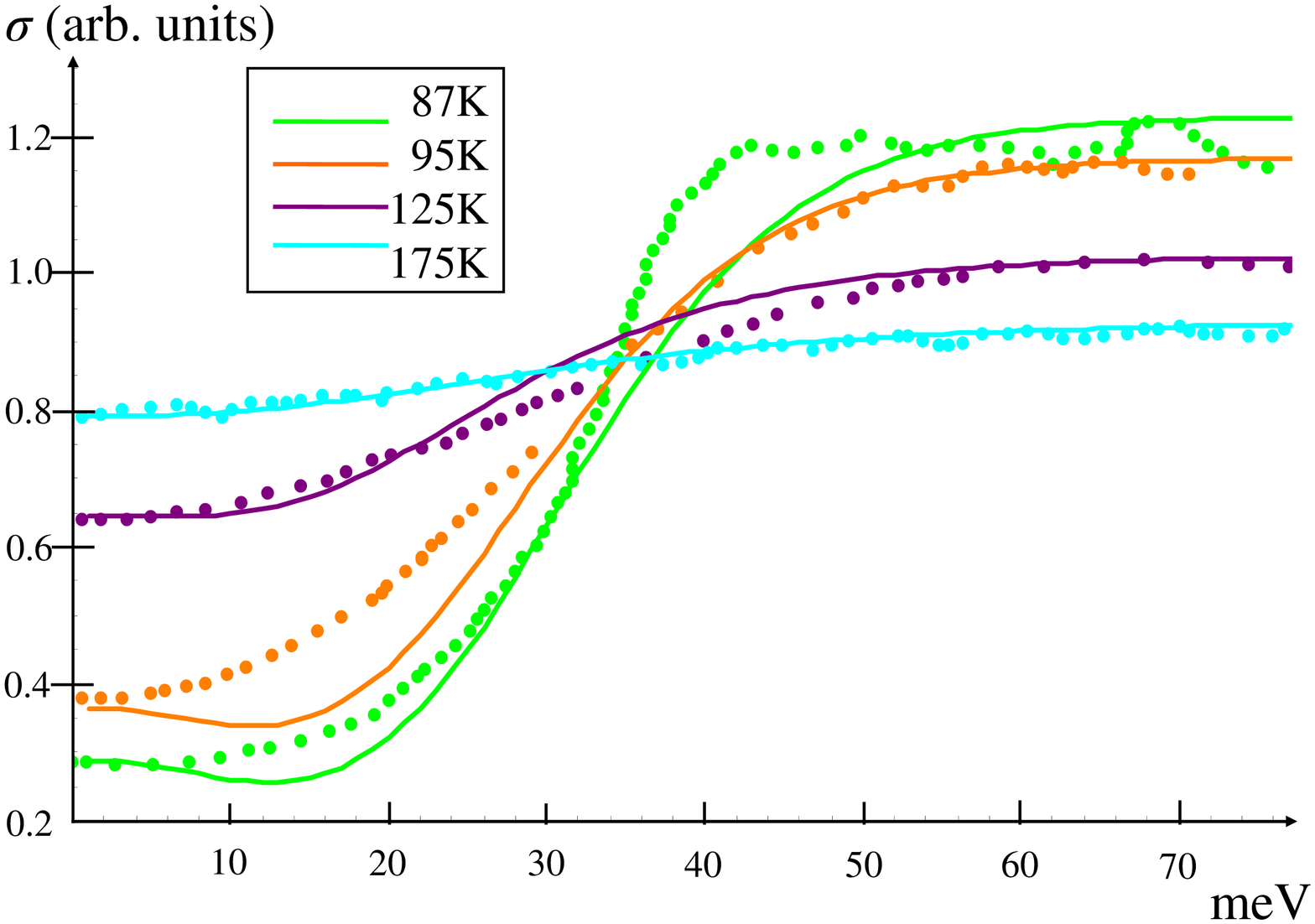}
 \caption{\small{(Colour online) Normal-state tunnelling conductance of a bosonic superconductor (solid lines) with fixed $\Gamma=10$meV and $Bx^2/P^2=1.96$ for each temperature, but changing $\Delta_p$ (see Fig.\ref{deltadependence} for temperature dependence of $\Delta_p$) compared to experimental conductance \cite{kras} for a few temperatures.}}\label{alltempsfit}
\end{center}
\end{figure}

Finally consider the normal-state, where $\Delta_{c\nu}=0$, which means $\epsilon_{\nu}=\sqrt{\xi_{\nu}^2+\Delta_{c\nu}^2}=\left|\xi_{\nu}\right|$, thus $u_{\nu}^2=1$ and $v_{\nu}^2=0$ for positive $\xi_{\nu}$, and $u_{\nu}^2=0$, $v_{\nu}^2=1$ for negative $\xi_{\nu}$. We find:
\begin{eqnarray}
I_{SS}(V)&=&\frac{2\pi e P^2}{\hbar} \sum_{\nu\nu^{\prime}}(f_{\nu}-f_{\nu^{\prime}})\delta(\xi_\nu+eV-\xi_{\nu^{\prime}}) \cr
& & + \frac{2\pi e B^2}{\hbar} \sum_{\nu \nu^{\prime}}\left[\left(1-f_\nu-f_{\nu^{\prime}}\right)\frac{x}{2}-f_{\nu}f_{\nu^{\prime}}\right] \left[\delta(\xi_\nu-eV+\xi_{\nu^{\prime}})-\delta (\xi_\nu+eV+\xi_{\nu^{\prime}})\right].\label{normalstate}
\end{eqnarray}
Now, we can follow the same steps as Ref.\cite{ourPRL} and neglect temperature effects by approximating the distribution function $f_{\nu}$, with a step function $\Theta(-\xi_{\nu})$ for temperatures near and above the transition temperature but sufficiently below the PG temperature $T^*=\frac{\Delta_p}{k_B}>T\gtrsim T_c$, and for high enough voltages, $\left|eV\right|\gtrsim k_BT$.
Using the model normal-state DOS, Eq.(\ref{modelDOS}) yields:
\begin{equation}
I_{SS}(V)\propto\frac{a^2}{2(a^2-1)}\left[\ln\frac{a^2(1+b^2)}{1+a^2b^2}-\frac{1}{a^2}\ln \frac{a^2+b^2}{1+b^2}\right]+\frac{B^2(1+x/2)}{P^2(a^2b^4-1)}\ln\frac{1+a^2b^2}{a(1+b^2)}+\frac{B^2xa^2}{2P^2(a^2-b^4)}\ln
\frac{a^2+b^2}{a(1+b^2)},
\end{equation}
where $a=e^{\frac{\left|eV\right|}{\Gamma}}$ and $b=e^{\frac{\Delta_p}{\Gamma}}$.
Near and above $T_c$, for high enough voltages $eV\gtrsim k_BT$, the conductance fits the experimental data and the gapped conductance is accounted for in underdoped mesas of Bi2212 \cite{kras}, as shown in Fig.\ref{normalstatessconductance}. For these voltages the shape of  the conductance is almost independent of the ratio $B/P $ and $x$ since the last term is much larger than all other terms in Eq.(24).
The gapped conductance at smaller voltages can be  accounted for by fully taking into account temperature effects in Eq.(\ref{normalstate}). One can show that the quadratic term $f_{\nu}f_{\nu^{\prime}}$ in Eq.(\ref{normalstate}) gives negligible contribution in the relevant temperature range, so differentiating the current with respect to the voltage yields the SS conductance as:
\begin{equation}
\sigma_{SS}\propto\int_{-\infty}^{\infty}d\xi\left[\text{sech}^2\left(\frac{\xi+eV-\Delta_p}{\Gamma}\right)+\text{sech}^2\left(\frac{\xi-eV-\Delta_p}{\Gamma}\right)\right]\left[\rho_N(\xi) f(\xi)+\frac{B^2x}{2P^2}\rho_N(-\xi)\left[f(\xi)-f(-\xi)\right]\right].
\end{equation}
This equation is plotted in Fig.\ref{alltempsfit} with  $\Gamma=10$meV, the ratio $B^2x/2P^2$ is fixed at 1.96 and we change $\Delta_p$ to fit the experimental temperature dependence. Our theoretical results closely resemble those found experimentally by Krasnov \cite{kras}.
Our comparison and other experiments \cite{parker2010, krasetal} suggests that the PG gradually  decreases with increasing temperature, which could be explained by many-particle effects at significant doping. We suggest that with increasing temperature thermally excited mobile polarons screen the EPI, so the binding energy of the bipolarons drops \cite{alekabmot} as with doping.
 \begin{figure}
\begin{center}
\includegraphics[angle=-90, width=0.6\textwidth]{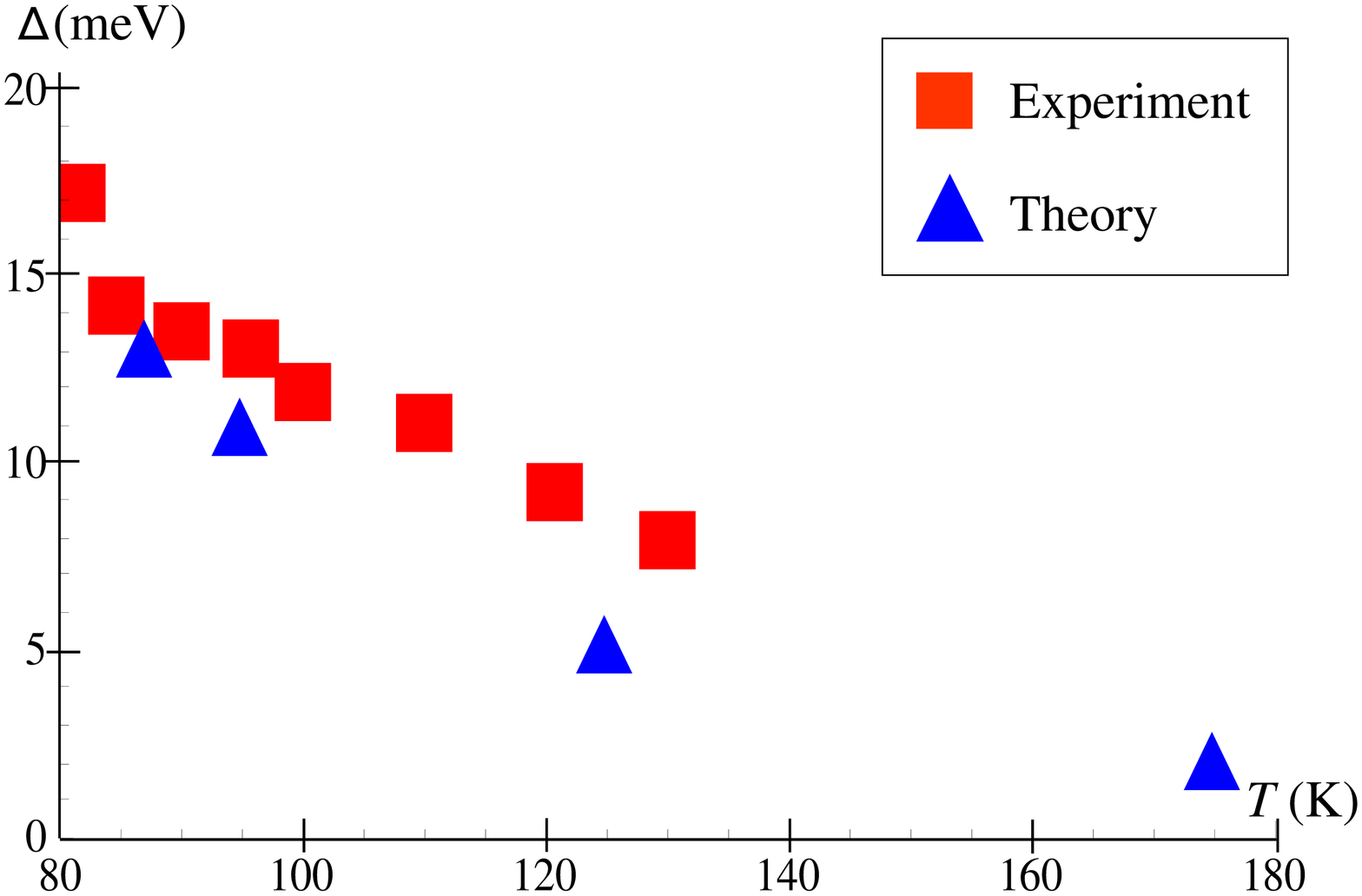}
 \caption{\small{(Colour online) Temperature dependence of $\Delta$ reflects what is expected from experimental results \cite{kras}.}}\label{deltadependence}
\end{center}
\end{figure}

Negative excess resistance below the transition temperature can be seen in cuprates \cite{krasnovPRL}. Our theory can account for this. Expanding Eqs.(\ref{equationss1}) and (\ref{normalstate}) in powers of $eV$ gives a zero bias conductance, for low temperatures in the superconducting state this is $\sigma_S (0) \propto T^{-1} \int_{0}^{\infty}d\epsilon\rho_S(\epsilon)^2\cosh(\epsilon/2k_BT)^{-2}$, and in the normal-state $\sigma_N (0) \propto T^{-1} \int_{-\infty}^{\infty}d\xi \rho_N(\xi)^2\cosh(\xi/2k_BT)^{-2}$. These integrals can be estimated to give respectively $\sigma_S (0)\propto T^{-1}\exp(-\Delta_c/k_BT)$ for the $s$-wave coherent gap, or $\sigma_S(0)\propto T^{2}$ for the $d$-wave gap, and $\sigma_N (0)\propto T^{-1}\exp(-T^*/T)$. The latter expression is in excellent agreement with the temperature dependence of the mesa tunnelling conductance above $T_c$ \cite{krasnovPRL} (see also Ref.\cite{alekabmot}). Extrapolating this expression to temperatures below $T_c$ yields the resistance ratio $R_S/R_N \propto e^{(\Delta_c/k_B-T^*)/T}$ ($s$-wave) or $R_S/R_N \propto\exp(-T^*/T)/T^2$ ($d$-wave). Hence in underdoped cuprates, where $T^*> \Delta_c/k_B$, the zero-bias tunnelling resistance at temperatures below $T_c$ is smaller than the normal state resistance extrapolated from above $T_c$ to the same temperatures (i.e. the negative excess resistance), as observed \cite{krasnovPRL}.
%
%
\section{Summary}

Data from tunnelling experiments is invaluable as it gives a huge insight into the low-energy excitations and thus the way high-temperature superconductors work.

NS tunnelling in cuprates has indicated two energy scales, the first is the SG that vanishes above $T_c$. The second is the PG which remains a mystery, there is no general consensus as to what it is or why it exists. STM with cuprates has also shown that the tunnelling conductance of charge carriers in one direction (say tip to sample) is different to the tunnelling conduction in the other direction, this gives asymmetry in the tunnelling spectra. Again, there is no general consensus as to why this is the case. Remarkably, the position of the tip on the cuprate gives different tunnelling results.

Intrinsic tunnelling and break junction experiments have also indicated a two-gap structure. Recent experiments have provided evidence that the PG is dependent on the temperature and decreases as the temperature increases above $T_c$. Another inexplicable feature of SS tunnelling has been the negative excess resistance, where the zero-bias tunnelling resistance that is extrapolated from above to below $T_c$ is larger than that measured in the superconducting state below $T_c$.

To have a successful theory, each of these puzzles should be accounted for in both NS and SS tunnelling.

Our theory is based on the ab initio ``LDA+GTB" band structures of charge transfer Mott-Hubbard insulators with the doping of impurities causing bandtails in the normal-state superconductor DOS. We make an extension to the BCS theory in the strong-coupling regime with bosonic (bipolaronic) carriers which are the real-space pairs of polarons. This theory has allowed us to describe the above unusual features of the cuprates.

We greatly appreciate valuable discussions with Ivan Bozovic, Jozef Devreese, Kenjiro Gomes, Ruihua He, Viktor Kabanov, Vladimir Krasnov, Nikolai Kristoffel, Klim Kugel and Dragan Mihailovic.




\end{document}